\documentclass[aps,prl,twocolumn,groupedaddress,superscriptaddress]{revtex4}

\usepackage{graphicx}
\usepackage{colortbl}
\makeatletter
\def\btt#1{\texttt{\@backslashchar#1}}%
\DeclareRobustCommand\bblash{\btt{\@backslashchar}}%
\makeatother

\begin{document}


\title{Emergence of non-equilibrium charge dynamics in a charge-cluster glass}

\author{T.~Sato}
\affiliation{Department of Applied Physics, University of Tokyo, Tokyo 113-8656, Japan}

\author{F.~Kagawa}
\email{fumitaka.kagawa@riken.jp}
\affiliation{Department of Applied Physics, University of Tokyo, Tokyo 113-8656, Japan}
\affiliation{RIKEN Center for Emergent Matter Science (CEMS), Wako 351-0198, Japan}

\author{K.~Kobayashi}
\affiliation{Condensed Matter Research Center (CMRC) and Photon Factory, Institute of Materials Structure Science, High Energy Accelerator Research Organization (KEK), Tsukuba 305-0801, Japan}

\author{K.~Miyagawa}
\affiliation{Department of Applied Physics, University of Tokyo, Tokyo 113-8656, Japan}

\author{K. Kanoda}
\affiliation{Department of Applied Physics, University of Tokyo, Tokyo 113-8656, Japan}

\author{R.~Kumai}
\affiliation{Condensed Matter Research Center (CMRC) and Photon Factory, Institute of Materials Structure Science, High Energy Accelerator Research Organization (KEK), Tsukuba 305-0801, Japan}

\author{Y.~Murakami}
\affiliation{Condensed Matter Research Center (CMRC) and Photon Factory, Institute of Materials Structure Science, High Energy Accelerator Research Organization (KEK), Tsukuba 305-0801, Japan}

\author{Y.~Tokura}
\affiliation{Department of Applied Physics, University of Tokyo, Tokyo 113-8656, Japan}
\affiliation{RIKEN Center for Emergent Matter Science (CEMS), Wako 351-0198, Japan}

\date{\today}

\begin{abstract}
Non-equilibrium charge dynamics, such as cooling-rate-dependent charge vitrification and physical aging, have been demonstrated for a charge-cluster glass in $\theta$-(BEDT-TTF)$_2$CsZn(SCN)$_4$ using electron transport measurements. The temperature evolution of the relaxation time obeys the Arrhenius law, indicating that the glass-forming charge liquid can be classified as a ``strong'' liquid in the scheme of canonical structural-glass formers. X-ray diffuse scattering further reveals that the spatial growth of charge clusters in the charge liquid is frozen below the glass transition temperature, indicating an intrinsic relationship between dynamics and structure in the charge-cluster glass. 
\end{abstract}


\maketitle

Glassy states can be found in various many-body systems when long-range ordering (or ``crystallization'') is avoided, for example, by randomness, rapid cooling, or competing interactions \cite{DebenedettiNature, TanakaReview, DagottoScience, SchmalianPRL}. Prototypical examples include structural glasses formed by certain supercooled liquids \cite{DebenedettiNature, Ediger, TanakaReview} and spin glasses in disordered magnets \cite{spinglassRMP, spinglass1, spinglass2}. However, quite recently, vitrification phenomena have been demonstrated in an organic charge-ordering system with a triangular lattice, $\theta$-(BEDT-TTF)$_2$RbZn(SCN)$_4$ (denoted by $\theta$-RbZn) [Fig.~1(a)] \cite{KagawaNatPhys}, although organic conductors are widely believed to be clean [BEDT-TTF (abbreviated as ET) denotes bis(ethylenedithio)tetrathiafulvalene]. When $\theta$-RbZn is rapidly cooled to avoid long-range charge ordering (LR-CO) [Fig.~1(b)], frozen short-range charge ordering (SR-CO) appears \cite{KagawaNatPhys}. Clearly, this spatially inhomogeneous state is neither a long-range ordering nor a ``para'' state, and we have therefore termed this state a charge-cluster glass to emphasize that the charge configuration is not completely random but is comprised of spatially heterogeneous regions \cite{KagawaNatPhys}. This is (most likely) a new class of glass-forming systems and clearly distinct from a glassy state in a structural degree of freedom; for example, in $\kappa$-ET salts, a glass transition in the ethylene-group conformation occurs irrespective of electronic ground states \cite{AkutsuPRB, MullerPRB}.

One of the hallmarks of glassy states is that they fall out of thermodynamic equilibrium of a liquid-like ``para'' state. Thus, close to the glass transition temperature $T_{\rm g}$, intriguing non-equilibrium phenomena appear, such as cooling-rate-dependent vitrification and physical aging \cite{AgingBook, LehenyPRB, LunkenheimerPRL}. However, the charge-cluster glass material $\theta$-RbZn exhibits a strong tendency towards a first-order structural transition that simultaneously stabilizes LR-CO (or ``charge crystallization'') [Fig.~1(b)] \cite{NogamiJPSJ}. This instability leads to difficulties in investigating non-equilibrium phenomena in $\theta$-RbZn while maintaining the charge-cluster glass; even below $T_{\rm g}$, a relaxation towards long-range charge crystallization inevitably overwhelms possible physical aging within the glass state. A clear demonstration of the thermodynamic non-equilibrium is of importance not only to provide an unequivocal evidence of the glassy state but also to explore the diversity and universality in the glass physics.

\begin{figure}
\includegraphics[width=8.3cm,clip]{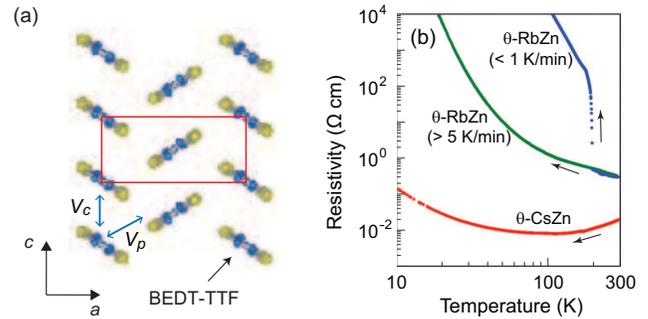}
\caption{(Color online) (a) Molecular arrangement in the ET conducting layer in $\theta$-(ET)$_2$$X$ [$X$ = RbZn(SCN)$_4$ and CsZn(SCN)$_4$] at room temperature: the rectangle represents the unit cell, and the 
two inequivalent inter-site Coulomb interactions \textit{$V_p$} and \textit{$V_c$} are shown by arrows.
(b) Resistivity temperature profiles obtained upon cooling for slowly cooled $\theta$-CsZn and $\theta$-RbZn and rapidly cooled $\theta$-RbZn, shown using a double logarithmic scale. In this scale, the cooling-rate dependence of $\theta$-CsZn is negligibly small in the range of 0.1-10 K/min.
}
\label{Fig1} 
\end{figure}

In this Letter, we investigate a $\theta$-RbZn analog, $\theta$-(ET)$_2$CsZn(SCN)$_4$ (denoted by $\theta$-CsZn), which does not exhibit LR-CO, at least at the laboratory time scale \cite{WatanabeJPSJ1999, SuzukiJPSJ, NadJPhys, ChibaPRB}. Cooling-rate-dependent charge vitrification and non-equilibrium aging behavior are successfully demonstrated using electron transport measurements, which have not been previously obtained for $\theta$-RbZn. Diffuse X-ray scattering measurements are further performed to investigate the temperature evolution of the charge clusters, revealing that the glassy charge dynamics are closely related to the spatial growth of the charge clusters.

The crystal structure of $\theta$-(ET)$_2$$X$ consists of alternating layers of conducting ET molecules and insulating anions \cite{HatsumiPRB}. The conduction band comprised of the highest occupied molecular orbital (HOMO) of the ET is hole-1/4-filled; thus, the charge liquid (i.e., the ``para'' state where the charges are delocalized) is subject to a charge-ordering instability due to inter-site Coulombic repulsion $V$ \cite{SeoJPSJ}. However, the frustration of $V$ associated with the triangular lattice [Fig.~1(a)] can potentially undermine the tendency towards LR-CO \cite{MerinoPRB}. In fact, charge ordering is avoided in $\theta$-RbZn by rapid cooling ($>$ 5 K/min) [Fig.~1(b)] \cite{KagawaNatPhys, NadPRB}. In contrast, $\theta$-CsZn does not exhibit LR-CO even upon slow cooling \cite{WatanabeJPSJ1999, SuzukiJPSJ, NadJPhys, ChibaPRB}. For instance, there is no signature of first-order charge ordering in the resistivity-temperature ($\rho$-$T$) profile [Fig.~1(b)]. Moreover, the existence of slow charge dynamics ($\sim$kHz) \cite{ChibaPRB} and SR-CO \cite{WatanabeJPSJ1999} has been confirmed at high temperatures. These experimental observations are reminiscent of the charge liquid state in $\theta$-RbZn \cite{ChibaPRL, WatanabeJPSJ2004}, suggesting that $\theta$-CsZn is a promising charge-cluster glass former. However, the evolution of charge vitrification has not yet been systematically studied for $\theta$-CsZn: the existence of $T_{\rm g}$ itself remains unclear and must first be resolved.

\begin{figure}
\includegraphics[width=8.3cm,clip]{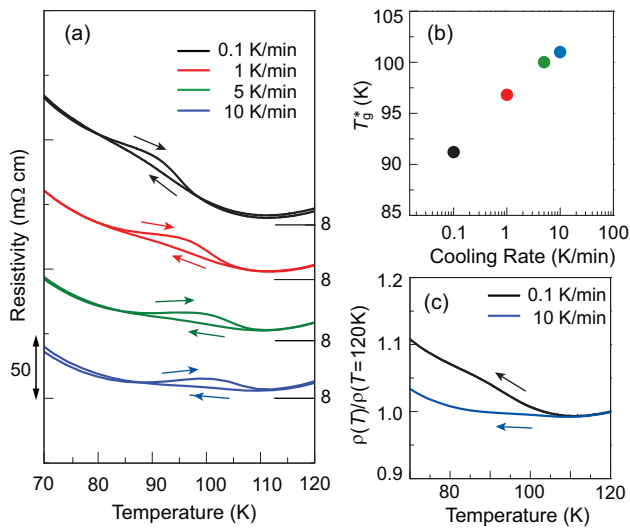}
\caption{(Color online) (a) Temperature dependence of resistivity at various temperature-sweeping rates. (b) Glass transition temperature versus temperature-sweeping rate. (c) Normalized resistivity profiles upon cooling at 0.1 K/min and 10 K/min.
}
\label{Fig2}
\end{figure}

In general, the glass transition is of kinetic origin and occurs when the time scale of the fluctuations exceeds the time available for molecular relaxation allowed by the cooling rate; hence, faster cooling leads to a higher experimental $T_{\rm g}^*$ \cite{BruningPRB}. To uncover this signature, we revisited the $\rho$-$T$ profile of $\theta$-CsZn. We note that appreciable hysteresis is present at approximately 90-100 K [Fig.~2(a)], which is consistent with a previous report \cite{NadJPhys}. It has been argued in the literature that this hysteresis is a signature of an inhomogeneously broadened first-order phase transition accompanied by a new charge configuration. In contrast, here we consider this hysteresis to be a manifestation of the formation of a charge-cluster glass.

To confirm our hypothesis, the thermal hysteresis was investigated under temperature-sweeping rates $Q$ of 0.1, 1, 5, and 10 K/min. The results are summarized in Fig.~\ref{Fig2}(a) and exhibit two noteworthy features. First, the hysteresis region depends on $Q$. For clarity, we (tentatively) define $T_{\rm g}^*$, which is determined from the $\rho$-$T$ profile, as the temperature at which the difference in the resistivity between the heating and cooling processes exhibits a maximum. Figure 2(b) clearly shows a systematic shift in $T_{\rm g}^*$ and a positive correlation between $T_{\rm g}^*$ and $Q$; therefore, the temperature hysteresis that accompanies $T_{\rm g}^*$ must be kinetic rather than thermodynamic in origin. Moreover, these experimental observations are in good agreement with general glass behavior \cite{BruningPRB}. Second, the resistivity below $T_{\rm g}^*$ is appreciably dependent on $Q$. Figure 2(c) shows that the data obtained at the 10-K/min cooling rate branch off from the data obtained at 0.1 K/min below $\sim$105 K. This deviation provides further evidence of the glass transition upon cooling at 10 K/min, and this temperature is reasonably close to $T_{\rm g}^*$ in Fig.~2(b) ($\approx$101 K for 10 K/min).

\begin{figure}
\includegraphics[width=8.3cm]{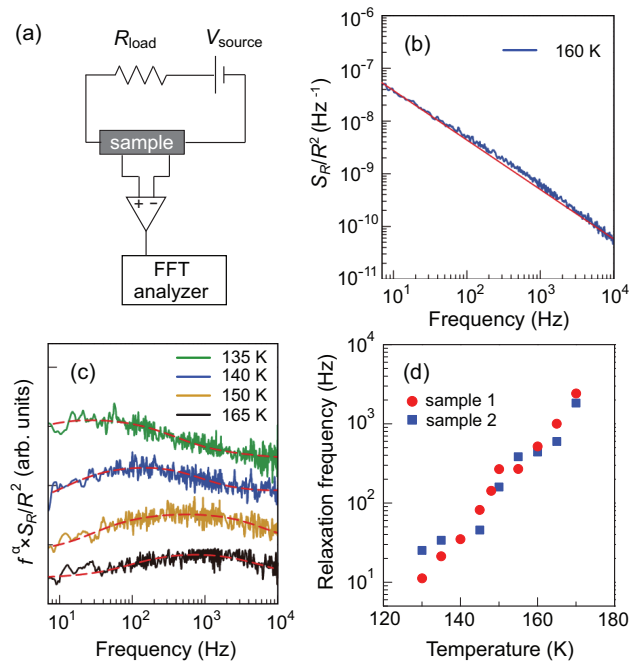}
\caption{(Color online) (a) Circuit for four-probe resistance fluctuation spectroscopy. (b) Typical power spectrum density of resistance fluctuations, $S_R$/$R^2$: the straight line is a fit to $1/f^\alpha$.
(c) Power spectrum density multiplied by frequency at various temperatures: broken curves are fits to continuously distributed Lorentzians plus $1/f^\alpha$ \cite{KagawaNatPhys}.
(d) Temperature dependence of the characteristic frequency extracted from the fits in (c).
}
\label{Fig3} 
\end{figure}

The existence of the charge-glass transition is further corroborated by observations of its precursor, i.e., slow charge dynamics, for thermodynamic equilibrium states above $T_{\rm g}^*$. Resistance fluctuation spectroscopy (i.e., noise measurements) is a powerful tool for measuring slow charge dynamics \cite{MullerChemPhysChem, KagawaNatPhys}. This method was applied to $\theta$-CsZn in this study. During the measurements, a steady current was applied along the $c$-axis such that the linear response was satisfied, and the voltage between the voltage-probing electrodes was fed into a FFT (Fast Fourier Transform) analyzer (Agilent 35670A) [Fig.~3(a)].

The typical noise power spectral density (PSD) of the resistance fluctuations \cite{MullerChemPhysChem}, $S_R/R^2$, is shown in Fig.~3(b). The global PSD is well-fitted by the so-called $1/f$ noise (the straight line), i.e., $1/f^\alpha$ with $\alpha$ $\sim$0.9-1.1 (corresponding to a weak temperature dependence) and $f$ denotes the frequency. The resistance fluctuations with a characteristic relaxation frequency, $f_0$, can be observed as a marginal deviation from the $1/f^\alpha$ fit [Fig.~3(b)]. For clarity, $f^\alpha \times S_R/R^2$ versus $f$ is plotted in Fig.~3(c); within this representation, resistance fluctuations with a frequency of $f_0$ appear as a broad peak rising out of a constant background. To extract $f_0$, we used a hypothetical superposition of continuously distributed Lorentzians with high-frequency $f_{\rm c1}$ and low-frequency $f_{\rm c2}$ cutoffs plus $1/f^\alpha$ \cite{KagawaNatPhys}. We found that the fitting curves (the broken curves) reproduced the spectra well [Fig.~3(c)], which enabled us to extract the temperature dependence of $f_0$ [$\equiv$ ($f_{\rm c1}$$f_{\rm c2}$)$^{1/2}$], as shown in Fig.~3(d). Here, a dramatic decrease in $f_0$ of several orders of magnitude was observed upon cooling for two different specimens [Fig.~3(d)], showing that the charge dynamics at equilibrium slow down as the glass transition is approached ($\approx$100 K). Moreover, as shown below, we confirmed that the relaxation time reaches 100-1000 s at $\approx$100 K, which is consistent with the conventional definition of $T_{\rm g}^*$ [see Fig.~4(b)].

\begin{figure}
\includegraphics[width=5.6cm]{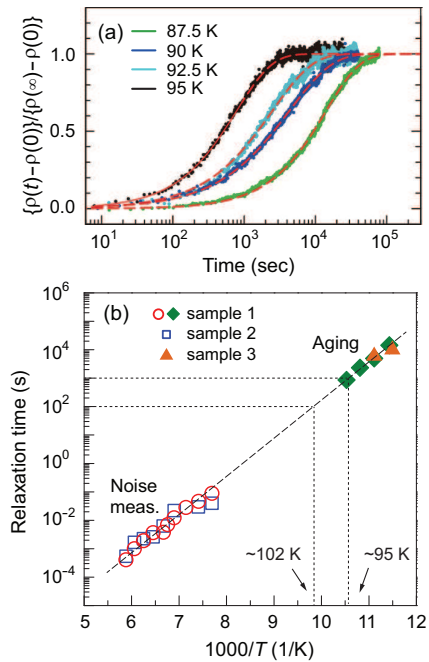}
\caption{(Color online) (a) Aging behavior of resistance as a function of time at various temperatures. The broken curves are fits to the phenomenological relaxation behavior, Eq. (1). (b) Temperature dependence of the relaxation time, $\tau_{\rm aging}$, derived from the fits in (a) (closed symbols). The results shown in Fig.~2(d) are also replotted in units of seconds (open symbols). The broken line is a fit to the Arrhenius behavior, $\propto$ exp($-$$\Delta$/$k_{\rm B}T$) with $\Delta$/$k_{\rm B}$ $\approx$ 2600 K. The dotted lines bound the temperature region where the relaxation time becomes 100-1000 s, which is a widely used definition of $T_{\rm g}^*$.}
\label{Fig4}
\end{figure}

Having established the existence of $T_{\rm g}^*$, it is reasonable to expect a non-equilibrium electronic state below $T_{\rm g}^*$. In fact, this signature is already apparent in Fig.~2(c): the low-temperature resistivity ($<$ $T_{\rm g}^*$$\approx$100 K) is obviously $Q$-dependent. Even more compelling evidence of non-equilibrium electronic states can be obtained by observing physical aging. For this purpose, the sample was first cooled down ($\sim$5 K/min) from high temperatures to a target temperature, and then, the time evolution of the resistance was recorded while the temperature was held fixed. Figure 4(a) shows the time ($t$) evolution of the resistivity at various temperatures. As expected, aging behavior clearly occurs for the resistivity below $T_{\rm g}^*$, indicating that the charge configurations are falling out of thermodynamic equilibrium with a very long relaxation time (e.g., up to several hours at 87.5 K).

The aging behavior can be further analyzed by using the well-known Kohlrausch-Williams-Watts (KWW) law, which is widely used to describe relaxation processes in supercooled liquids \cite{EdigerJPhysChem, AngellJAP}: 
\begin{equation}
\rho(t) = \rho_0 + (\rho_\infty-\rho_0) [1-\exp \{-(t/\tau_{\rm aging})^\beta \} ],
\end{equation}
where $\rho_0$ and $\rho_\infty$ denote the initial and final resistivity values during the aging process, respectively, and $\tau_{\rm aging}$ and $\beta$ denote the relaxation time and the stretching parameter, respectively. Figure 4(b) displays the temperature dependence of $\tau_{\rm aging}$ in the Arrhenius representation together with the relaxation times, $\tau_{\rm noise}$ ($\equiv$ 1/$f_0$), that were extracted using resistance fluctuation spectroscopy above $T_{\rm g}^*$ [Fig.~3(d)]. Remarkably, the temperature profiles of $\tau_{\rm aging}$ and $\tau_{\rm noise}$ appear to obey the same equation, $\tau$ $\propto$ exp($-$$\Delta$/$k_{\rm B}T$) with $\Delta$/$k_{\rm B}$ $\approx$ 2600 K, suggesting common charge dynamics for the equilibrium states above $T_{\rm g}^*$ and the non-equilibrium states below $T_{\rm g}^*$ \cite{LunkenheimerPRL}. It is plausible to assume that the relevant charge dynamics consist of a rearrangement of the charge configurations, which may be accompanied by a distortion of the local lattice/molecules. Above $T_{\rm g}^*$, thermodynamic equilibrium is established instantaneously because $\tau$ is short, and consequently, the charge fluctuations are centered around the equilibrium states. In contrast, below $T_{\rm g}^*$, laboratory timescales (or greater) are required to reach thermodynamic equilibrium, and therefore, the relaxation process from an initial to a (quasi-)equilibrium final state is observed immediately after the temperature is changed: this process is precisely equivalent to aging.

The observed Arrhenius behavior has implications for the nature of the slow charge dynamics. Generally, the temperature dependence of the relaxation time in various glass systems can be of either an Arrhenius or Vogel-Fulcher-Tammann (VFT) type \cite{DebenedettiNature}; for instance, ``strong'' liquids, such as SiO$_2$, exhibit Arrhenius behavior, whereas ``fragile'' liquids, such as $o$-terphenyl, follow the VFT equation, $\propto$ exp$\{A/(T-T_0)\}$, where $A$ and $T_0$ are temperature-independent constants. Such super-Arrhenius behavior is often interpreted as the consequence of an increasing number of dynamically correlated molecules \cite{BauerPRL}. Within this scheme, the glass-forming charge liquid in $\theta$-CsZn can obviously be classified as a strong liquid. The general implications of Arrhenius behavior are that the glassy dynamics are dominated by an elementary process rather than a cooperative process: for instance, local breaking and reforming of Si-O bonds are considered to play a major role in the glassy dynamics of SiO$_2$. The strong-liquid nature of $\theta$-CsZn thus suggests that the rearrangement of charge configurations occurs locally. This implication appears to be compatible with the strong geometrical frustration in $\theta$-CsZn \cite{MoriJPSJ} because this frustration is expected to produce locally different configurations of similar energies, resulting in less cooperative dynamics.

\begin{figure}
\includegraphics[width=8.3cm]{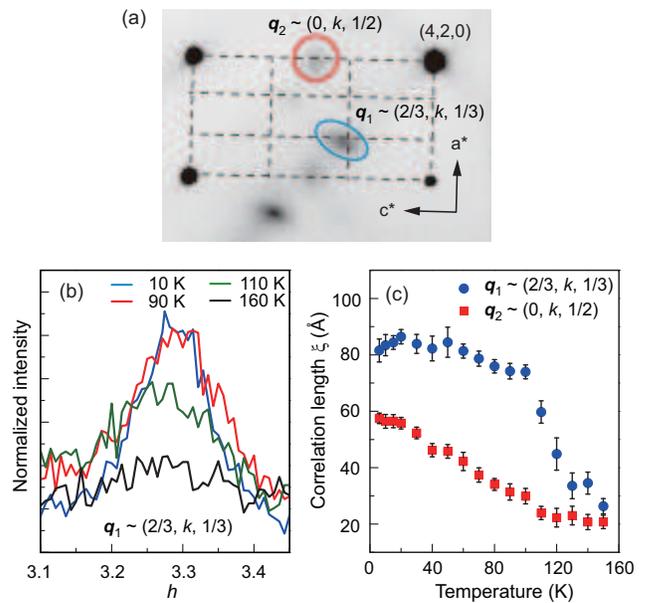}
\caption{(Color online) (a) Oscillation photograph of the $a^\ast$-$c^\ast$ plane at 33 K. Diffuse scattering patterns characterized by {\boldmath $q_1$} $\sim$$(2/3, k, 1/3)$ and {\boldmath $q_2$} $\sim$$(0, k, 1/2)$ can be observed and are indicated by an ellipsoid and a circle in the image, respectively.
(b) Line profile of the {\boldmath $q_1$} diffuse scattering patterns along the $a^\ast$ axis at various temperatures.
(c) Temperature dependence of the sizes of {\boldmath $q_1$} (circles) and {\boldmath $q_2$} (squares) charge clusters.
} 
\label{Fig5} 
\end{figure}

Finally, to obtain insights into how the development of slow dynamics correlates with the pre-existing SR-CO (or charge cluster), we conducted X-ray diffuse scattering measurements at a synchrotron facility \cite{KEK}. As previously reported \cite{WatanabeJPSJ1999, SawanoNature}, both ``3$\times$3''-period and ``1$\times$2''-period charge clusters were observed that could be characterized by the wave vectors {\boldmath $q_1$} $\sim$$(2/3, k, 1/3)$ and {\boldmath $q_2$} $\sim$(0, $k$, 1/2), respectively [Fig.~5(a)] ($k$ denotes negligible coherence between the ET layers).

In this study, it was found that the linewidth of the {\boldmath $q_1$} and {\boldmath $q_2$} charge clusters narrows (and hence the cluster size increases) at low temperatures [Fig.~5(b)]. To quantify this effect, the size of the charge clusters, $\xi$, was estimated as the inverse of the full-width at half-maximum of the line profiles. The $\xi$-$T$ profile for each of the clusters is shown in Fig.~5(c), where the size of the ``3$\times$3''-period charge clusters ({\boldmath $q_1$}) grows as the temperature decreases but levels off below $\sim$100 K; that is, the ``3$\times$3''-period charge clusters appear to be frozen. Remarkably, this temperature roughly coincides with the $T_{\rm g}^*$ value that was determined from charge dynamics considerations [Figs.~2(b) and 4(b)], indicating that the spatial growth of the ``3$\times$3''-period charge clusters is strongly coupled with the underlying, glassy charge dynamics. This result is also consistent with the charge vitrification reported in $\theta$-RbZn, in which only ``3$\times$4''-period charge clusters are observed that then freeze upon rapid cooling \cite{KagawaNatPhys}.

Interestingly, the ``1$\times$2''-period charge clusters ({\boldmath $q_2$}) continue to grow with decreasing temperatures even below $T_{\rm g}^*$ ($\approx$100 K), suggesting that the charge dynamics associated with the ``1$\times$2''-period charge clusters are still active. Conversely, there appears to be ``free space'' for the ``1$\times$2''-period charge clusters to grow. Based on these observations, we conjecture that at approximately $T_{\rm g}^*$, sparsely frozen, ``3$\times$3''-period charge clusters exist in the sea of the ``para'' state, thereby allowing subsequent growth of another type of charge clusters at lower temperatures.

In conclusion, we have demonstrated the emergence of non-equilibrium charge dynamics from a charge-cluster glass in $\theta$-CsZn by observing cooling-rate-dependent charge vitrification and physical aging of the resistance. The temperature dependence of the relaxation time follows the Arrhenius law, suggesting that the glassy charge dynamics are caused by the local rearrangement of charge configurations. Moreover, X-ray diffuse scattering measurements revealed that the spatial growth of the ``3$\times$3''-period charge clusters is closely related to the glassy charge dynamics. All of these experimental observations show that the strongly correlated electron systems in the $\theta$-(ET)$_2X$ family are unequivocally charge-glass formers.

The authors thank H. Tanaka and H. Seo for fruitful discussions. This work was performed under the approval of the Photon Factory Program Advisory Committee (Proposal No. 2012G115). This work was partially supported by the Funding Program of World-Leading Innovative R$\&$D on Science and Technology (FIRST program) for Quantum Science on Strong Correlation initiated by the Council for Science and Technology Policy, Japan, and by JSPS KAKENHI (Grant Nos. 24224009, 24684020, 23340111, 24654101 and 25220709).


\begin{references}

\bibitem{DebenedettiNature} P. G. Debenedetti and F. H. Stillinger,
Nature \textbf{410}, 259 (2001).


\bibitem{TanakaReview} H. Tanaka,
Eur. Phys. J. E \textbf{35}, 113 (2012).

\bibitem{SchmalianPRL} J. Schmalian and P. G. Wolynes,
Phys. Rev. Lett. \textbf{85}, 836 (2000).

\bibitem{DagottoScience} E. Dagotto,
Science \textbf{309}, 257 (2005).

\bibitem{Ediger} M. D. Ediger, 
Annu. Rev. Phys. Chem. \textbf{51}, 99 (2000).

\bibitem{spinglassRMP} A. P. Binder and K. Young, 
Rev. Mod. Phys. \textbf {58}, 801 (1986).

\bibitem{spinglass1} R. V. Chamberlin, 
Phys. Rev. B \textbf{30}, 5393 (1984). 

\bibitem{spinglass2} L. Lundgren, P. Svedlindh, P. Nordblad, and O. Beckman, 
Phys. Rev. Lett. \textbf {51}, 911 (1983). 

\bibitem{KagawaNatPhys} F. Kagawa, T. Sato, K. Miyagawa, K. Kanoda, Y. Tokura, K. Kobayashi, R. Kumai, and Y. Murakami,
Nature Phys. \textbf{9}, 419 (2013).


\bibitem{AkutsuPRB} H. Akutsu, K. Saito, and M. Sorai,
Phys. Rev. B \textbf{61}, 4346 (2000).

\bibitem{MullerPRB} J. M$\rm {\ddot{u}}$ller, M. Lang, F. Steglich, J. A. Schlueter, A. M. Kini, and T. Sasaki,
Phys. Rev. B \textbf{65}, 144521 (2002).

\bibitem{AgingBook} L. C. E. Struik, \textit{Physical Aging in Amorphous Polymers and Other Materials} (Elsevier, Amsterdam, 1978).

\bibitem{LehenyPRB} R. L. Leheny and S. R. Nagel, 
Phys. Rev. B \textbf{57}, 5154 (1998).

\bibitem{LunkenheimerPRL} P. Lunkenheimer, R. Wehn, U. Schneider, and A. Loidl,
Phys. Rev. Lett. \textbf{95}, 055702 (2005).


\bibitem{NogamiJPSJ} Y. Nogami, N. Hanasaki, M. Watanabe, K. Yamamoto, T. Ito, N. Ikeda, H. Ohsumi, H. Toyokawa, Y. Noda, I. Terasaki, H. Mori, and T. Mori, 
J. Phys. Soc. Jpn. \textbf{79}, 044606 (2010).

\bibitem{WatanabeJPSJ1999} M. Watanabe, Y. Nogami, K. Oshima, H. Mori, and S. Tanaka,
J. Phys. Soc. Jpn. \textbf{68}, 2654 (1999).

\bibitem{SuzukiJPSJ} K. Suzuki, K. Yamamoto, K. Yakushi, and A. Kawamoto,
J. Phys. Soc. Jpn. \textbf{74}, 2631 (2005).

\bibitem{NadJPhys} F. Nad, P. Monceau, and H. M. Yamamoto,
J. Phys.: Condens. Matter. \textbf{20}, 485211 (2008).

\bibitem{ChibaPRB} R. Chiba, K. Hiraki, T. Takahashi, H. M. Yamamoto, and T. Nakamura,
Phys. Rev. B \textbf{77}, 115113 (2008).


\bibitem{HatsumiPRB} H. Mori, S. Tanaka, and T. Mori, 
Phys. Rev. B \textbf{57}, 12023 (1998).


\bibitem{SeoJPSJ} H. Seo, 
J. Phys. Soc. Jpn \textbf{69}, 805 (2000).

\bibitem{MerinoPRB} J. Merino, H. Seo, and M. Ogata,
Phys. Rev. B \textbf{71}, 125111 (2005).

\bibitem{NadPRB} F. Nad, P. Monceau, and H. M. Yamamoto,
Phys. Rev. B \textbf{76}, 205101 (2007).



\bibitem{ChibaPRL} R. Chiba, K. Hiraki, T. Takahashi, H. M. Yamamoto, and T. Nakamura, 
Phys. Rev. Lett. \textbf{93}, 216405 (2004).

\bibitem{WatanabeJPSJ2004} M. Watanabe, Y. Noda, Y. Nogami, and H. Mori,
J. Phys. Soc. Jpn. \textbf{73}, 116 (2004).

\bibitem{BruningPRB} R. Br$\rm {\ddot{u}}$ning, and K. Samwer, 
Phys. Rev. B \textbf{46}, 11318 (1992).

\bibitem{MullerChemPhysChem} J. M$\rm {\ddot{u}}$ller,
ChemPhysChem. \textbf{12}, 1222 (2011).


\bibitem{EdigerJPhysChem} M. D. Ediger, C. A. Angell, and S. R. Nagel, 
J. Phys.Chem. \textbf{100}, 13200 (1996).

\bibitem{AngellJAP} C. A. Angell, K. L. Ngai, G. B. McKenna, P. F. McMillan, and S. W. Martin,
J. Appl. Phys. \textbf{88}, 3113 (2000).

\bibitem{BauerPRL} Th. Bauer, P. Lunkenheimer, and A. Loidl,
Phys. Rev. Lett. \textbf{111}, 225702 (2013).

\bibitem{MoriJPSJ} T. Mori, 
J. Phys. Soc. Jpn. \textbf{72}, 1469 (2003).

\bibitem{KEK} BL-8A beamline of the photon Factory at KEK.

\bibitem{SawanoNature} F. Sawano, I. Terasaki, H. Mori, M. Watanabe, N. Ikeda, Y. Nogami, and Y. Noda,
Nature \textbf{437}, 522 (2005).



\end{references}
\end{document}